\documentclass[12pt,preprint]{aastex}
\usepackage{natbib}
\usepackage{graphicx}

\shorttitle{Stellar Mass Indicators}
\shortauthors{Lester, Dinshaw \& Neilson}

\begin{document}
\bibliographystyle{apj} % style file apj.bst

\title{Indicators of Mass in Spherical Stellar Atmospheres}

\author{John B. Lester}
\affil{Department of Chemical and Physical Sciences,
       University of Toronto Mississauga,
       Mississauga, Ont, L5L 1C6 Canada}
\and
\affil{Department of Astronomy \& Astrophysics, University of Toronto}
\email{lester@astro.utoronto.ca}

\author{Rayomond Dinshaw}
\affil{Department of Chemistry,
       University of Toronto,
       Toronto, Ont, M5S 3H6 Canada}
\email{rdinshaw@chem.utoronto.ca}
\and
\author{Hilding R. Neilson}
\affil{Department of Physics \& Astronomy,
       East Tennessee State University,
       Box 70652,
       Johnson City, TN, 37614, USA}
\email{neilsonh@etsu.edu}

\begin{abstract}
Mass is the most important stellar parameter, but it is not directly 
observable for a single star.  Spherical model stellar atmospheres are 
explicitly characterized by their luminosity ($L_\star$), mass 
($M_\star$) and radius ($R_\star$), and observations can now determine 
directly $L_\star$ and $R_\star$.  We computed spherical model 
atmospheres for red giants and for red supergiants holding $L_\star$ 
and $R_\star$ constant at characteristic values for each type of star 
but varying $M_\star$, and we searched the predicted flux spectra and 
surface-brightness distributions for features that changed with mass.  
For both stellar classes we found similar signatures of the star's mass 
in both the surface-brightness distribution and the flux spectrum.  The 
spectral features have been use previously to determine $\log_{10} (g)$,
and now that the luminosity and radius of a non-binary red giant or red 
supergiant can be observed, spherical model stellar atmospheres can be 
used to determine the star's mass from currently achievable 
spectroscopy.  The surface-brightness variations with mass are slightly 
smaller than can be resolved by current stellar imaging, but they offer 
the advantage of being less sensitive to the detailed chemical 
composition of the atmosphere.
\end{abstract}


\section{Introduction} \label{sec:sec1}
Plane-parallel model stellar atmospheres are characterized by their 
effective temperature, defined as 
\begin{equation} \label{eq:teff}
T_\mathrm{eff} \equiv \left (\frac{L_\star}
                                  {\sigma 4 \pi R^2_\star} 
                      \right )^{1/4},
\end{equation}
and surface gravity,
\begin{equation} \label{eq:logg}
g \equiv G \frac{M_\star}{R^2_\star}.
\end{equation}
As is apparent, these two parameters are degenerate expressions of the 
three more fundamental stellar parameters $L_\star$, $M_\star$ and 
$R_\star$.  Along the main sequence the variation of $R_\star$ is 
small, making $T_\mathrm{eff}$ and $\log_{10} (g)$ nearly unique 
identifiers of the star.  However, in the cool, luminous quadrant of 
the Hertzsprung-Russell diagram, stars with a large range of $R_\star$ 
and $M_\star$ converge to similar values of $T_\mathrm{eff}$ and 
$\log_{10} (g)$, which makes it difficult to establish the fundamental 
parameters of non-binary red giants and red supergiants.

Fortunately, optical/infrared interferometry is helping to break this 
degeneracy by measuring directly the angular diameters of stars.  By 
combining these angular diameters with stellar parallaxes that are 
being measured with increasing precision to ever greater distances, it 
is now possible to determine $R_\star$ observationally.  In addition, 
combining the stellar flux measured at Earth with the same stellar 
angular diameter and parallax yields the $L_\star$, assuming the star 
radiates isotropically.  With measured values of $L_\star$ and 
$R_\star$, the observational attention turns to $M_\star$, the key 
parameter that determines the other two.

It is now routine \citep{Hauschildt1999, Gustafsson2008, Lester2008} to 
compute spherically symmetric models of stellar atmospheres that are 
characterized explicitly by $L_\star$, $M_\star$ and $R_\star$, in 
addition to the chemical composition.  Assuming that red giants and 
supergiants are spherically symmetric, the ability to measure $L_\star$ 
and $R_\star$ raises the question whether spherical stellar atmospheres 
predict any indicators of stellar mass that could be determined 
observationally.

\section{Red Giants} \label{sec:sec2}

To search for possible indicators of stellar mass, the \textsc{SAtlas} 
program \citep{Lester2008} was used to compute a spherical model 
atmosphere with properties representative of a typical red giant: 
$L_\star = 500~L_{\sun}$, $M_\star = 0.8~M_{\sun}$ and 
$R_\star = 50~R_{\sun}$.  
The model was calculated with the opacity distribution function (ODF) 
version of \textsc{SAtlas} using the ODFs recomputed by Castelli 
(wwwuser.oat.ts.astro.it/castelli/) in 2012 February to incorporate 
Kurucz's corrections to his $\mathrm{H_2O}$ line list.  As Kurucz 
explains (kurucz.harvard.edu/molecules/H2O/h2ofastfix.readme), he uses 
the $gf$ values from \citet{Partridge1997} after renormalizing them to 
agree with the $gf$ values from the HITRAN (HIgh-resolution 
TRANsmission) molecular absorption database 
(www.cfa.harvard.edu/hitran/), but ignoring nuclear spin factors.  
Kurucz's original $\mathrm{H_2O}$ line list did not take into account 
that the $\mathrm{H_2^{16}O, H_2^{17}O, H_2^{18}O}$ and 
$\mathrm{HD^{16}O}$ isotopologs each has its own partition function.  
To correct for this, while also using only the $\mathrm{H_2^{16}O}$ 
partition function in \textsc{Atlas}, Kurucz assumes that, at stellar 
temperatures, the partition function of each isotopolog differs from 
the partition function of $\mathrm{H_2^{16}O}$ by a constant 
multiplicative factor, which he determines by comparing the $gf$ values 
of a few strong lines in the HITRAN database with the 
\citet{Partridge1997} values.
The ODF assumes a depth-independent microturbulent velocity of 2 km/s 
and
the solar abundances tabulated by \citet{Grevesse1998}.  The mean 
difference between the \citet{Grevesse1998} abundances and the more 
recent tabulation of \citet{Asplund2009} for the elements with number 
densities $\geq 10^{-10} \times N_\mathrm{H}$ is only 0.02 in 
$\log_{10}$ of the abundances.  The ODF also uses the large wavelength 
spacing with an average spectral resolving power of 
$R \equiv \lambda/\Delta \lambda\approx 50$, corresponding to 
$\Delta \lambda \approx 10$~nm at 500~nm.  A test using an ODF with the 
smaller wavelength space, which gives $\approx 5 \times$ higher 
spectral resolving power, produced a model structure (temperature as a 
function of gas pressure) that was almost indistinguishable from the 
structure of the model with the lower resolution ODF.  Therefore,  the 
lower resolution ODF was used to calculate all the models.
The model was iterated until the luminosity was constant 
to $\leq 1\%$ for $\log_{10} (\tau_\mathrm{Ross}) \geq 10^{-7}$ and the 
derivative of the luminosity was constant to $\leq 1\%$ for 
$\log_{10} (\tau_\mathrm{Ross}) \geq 10^{-3}$, where 
$\tau_\mathrm{Ross}$ is the Rosseland optical depth.

To test the dependence on mass, two additional red giant models were 
computed holding the values of $L_\star$ and $R_\star$ constant, but 
with $M_\star = 1.6~M_{\sun}$ and $M_\star = 2.4~M_{\sun}$.  These 
parameters correspond to $T_\mathrm{eff} = 3865$ K for all three stars, 
but $\log_{10} (g) = 0.94$ for the $0.8~M_{\sun}$ star, 
$\log_{10} (g) = 1.24$ for the $1.6~M_{\sun}$ star and 
$\log_{10} (g) = 1.42$ for the $2.4~M_{\sun}$ star.  
The effect of mass on the stellar atmosphere can be represented by a 
relative extension parameter, $\epsilon$, defined as the radial 
distance from the deepest visible layer of the atmosphere out to an 
arbitrary upper radius, normalized by the stellar radius, 
\begin{equation} \label{eq:eps}
\epsilon \equiv \frac{R(\mathrm{upper})-R(\mathrm{deep})}{R_\star}.
\end{equation}
A convenient value for $R(\mathrm{deep})$ is where 
$\tau_\mathrm{Ross} = 1$, and the radiation's probability of escape is 
$\exp(-1) = 0.36$.  The definition of the upper radius is arbitrary.  
Defining $R(\mathrm{upper})$ where $\tau_\mathrm{Ross} = -3$, 
corresponding to about three pressure scale heights, leads to 
$\epsilon(-3)$.  Using this, the $0.8~M_{\sun}$ star has 
$\epsilon(-3) \approx 0.025$, the $1.6~M_{\sun}$ star has 
$\epsilon(-3) \approx 0.012$ and the $2.4~M_{\sun}$ star has 
$\epsilon(-3) \approx 0.008$. Defining the upper $R$ is where 
$\tau_\mathrm{Ross} = -6$, leads to the definition of $\epsilon(-6)$.  
Using this, the $0.8~M_{\sun}$ star has 
$\epsilon(-6) \approx 0.06$, the $1.6~M_{\sun}$ star has 
$\epsilon(-6) \approx 0.028$ and the $2.4~M_{\sun}$ star has 
$\epsilon(-6) \approx 0.019$.  Clearly the extension parameter is 
greater with this choice of $R(\mathrm{upper})$, but by either 
definition these test models have modest extension.

\subsection{Surface Brightness Distribution} \label{subsec:sec2_1}

\begin{figure}
  \plotone{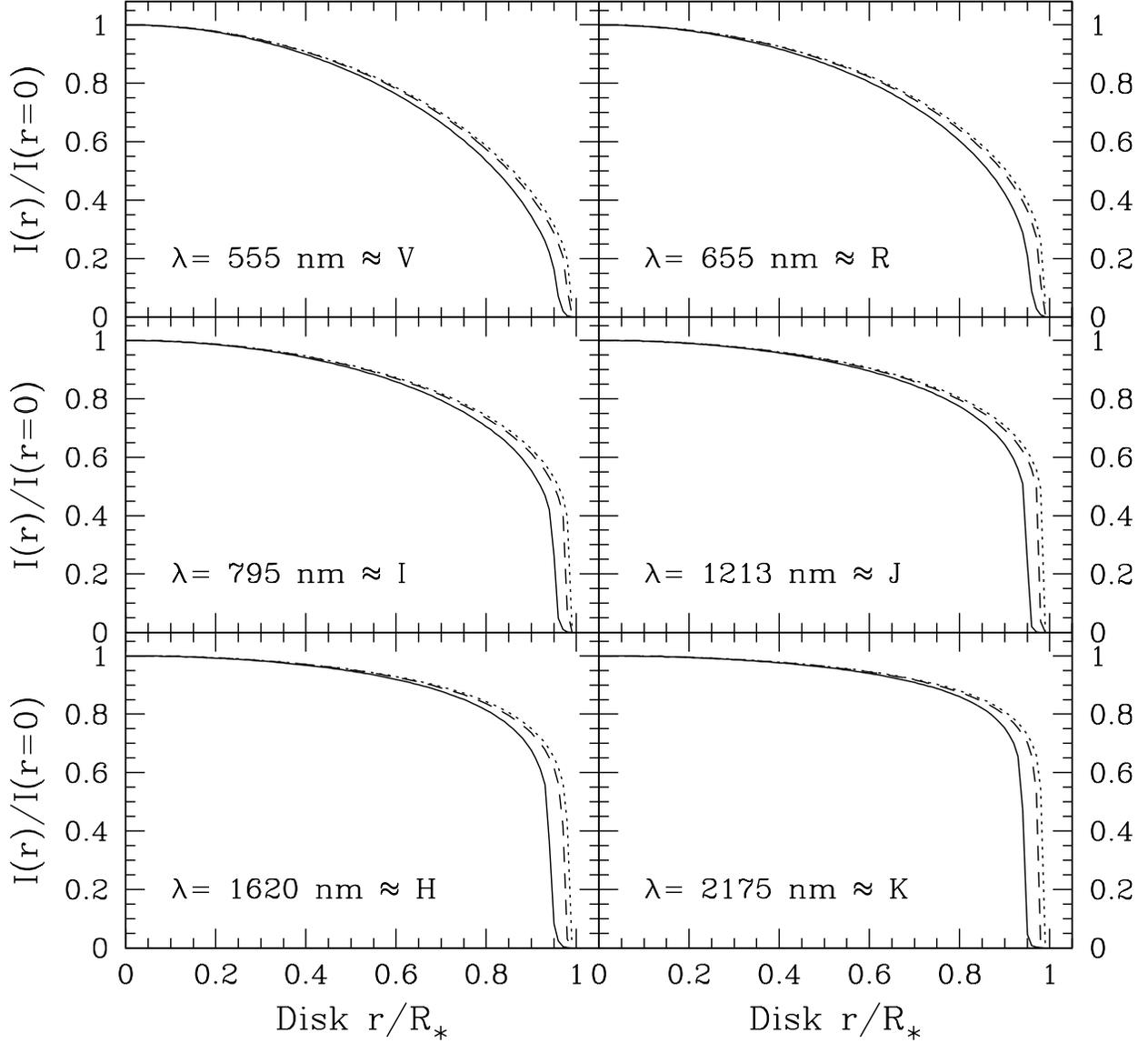}
  \caption{Surface brightness distributions for three red giant 
           spherical model atmospheres at several wavelengths.  The 
           atmospheres all have $L_\star = 500~L_{\sun}$ and 
           $R_\star = 50~R_{\sun}$.  The solid lines are for 
           $M_\star = 0.8~M_{\sun}$, the dashed lines are for 
           $M_\star = 1.6~M_{\sun}$ and the dotted lines are for 
           $M_\star = 2.4~M_{\sun}$.  The wavelengths are approximately 
           the effective wavelengths of the $V$, $R$, $I$, $J$, $K$ and 
           $H$ filter systems.}
  \label{fig:ldg6}
\end{figure}

The three red giant model atmospheres were used to calculate the 
surface intensity at 100 values of the disk's fractional radius, 
$r/R_\star$, ranging from 0 to 1 in steps of 0.01, for each wavelength 
bin of the ODF that had a non-negligible amount of radiation.  
A comparison was made of the surface brightness distributions of the 
same model parameters computed with both the lower and higher spectral 
resolution ODFs described in the previous section.  At wavelengths 
$\gtrsim 1000$~nm, the surface brightness distributions were almost 
identical for the two spectral resolutions because there are so few 
spectral lines.  At wavelengths $< 1000$~nm, the surface brightness 
distributions do differ very slightly because there are more and 
stronger spectral lines.  The ODF with the finer wavelength spacing 
resolves some spectral detail that is averaged out in the lower 
resolution ODF.  As a result, at some wavelengths the surface brightness
distribution is shifted slightly toward or away from $R_\star$, 
depending on whether the higher resolution ODF happens to include a 
strong line or not.  The lower spectral resolution surface brightness 
distribution are used here because they seem to be a better match to the
observations, but the synthetic spectra described in 
$\S$~\ref{subsec:sec2_2} can compute the surface brightness distribution
for any spectral resolving power, which can match any particular 
observation exactly.
Figure~\ref{fig:ldg6} shows the limb darkening curves for the three 
red giant models at the ODF wavelengths close to the effective 
wavelengths of the $V$ and $R$ magnitudes listed by \citet{Bessell2005} 
and to the effective wavelengths of the $I$, $J$, $H$ and $K$ 
magnitudes given by \citet{Tokunaga2000}.  It is apparent that 
increasing the stellar mass makes the relative intensity greater at 
every location on the stellar disk for every wavelength.  This trend 
results from the gas density and the thermal radiation of the 
atmosphere increasing with the stellar mass.  It is also apparent that 
the curves flatten as the wavelength increases toward the peak of the 
Planck distribution for the star's temperature.

The differences between the curves in Figure~\ref{fig:ldg6} uses a 
knowledge of the stellar limb's absolute location, $R_\star$, which is 
present in the models.  However, as the stellar  mass is reduced, the 
upper atmosphere becomes more tenuous, and the radiation from these 
outer radii decreases to the point where some of the curves in 
Figure~\ref{fig:ldg6} appear to go to zero before reaching $R_\star$.  
In fact, the relative intensity is not zero, but it is less than 0.01, 
making it too small to be resolved in the plot.  This reduced relative 
intensity at the stellar limb is a direct consequence of the spherical 
geometry of the models, and it also affects the observations.

\begin{figure}
  \plotone{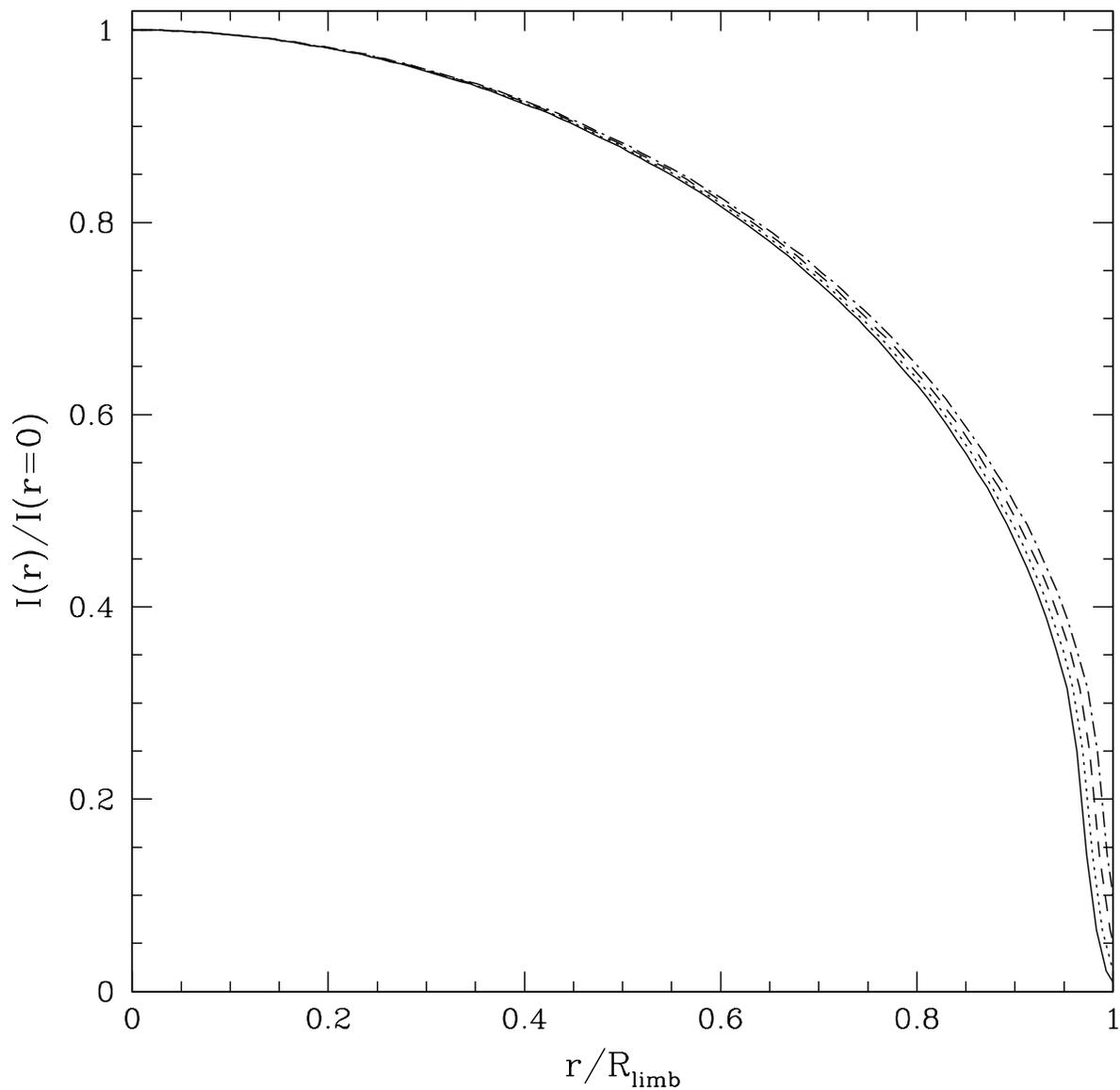}
  \caption{Renormalization of the limb-darkening profile at 
           $\lambda = 705$~nm for the spherical model atmosphere with 
           $L_\star = 500~L_{\sun}$, $R_\star = 50~R_{\sun}$ and 
           $M_\star = 0.8~M_{\sun}$.  The different curves represent 
           different choices for the value of ``$f$'' in 
           Equation~\ref{eq:rlimb} to define $R_\mathrm{limb}$. The 
           solid line is for $f = 0.01$, the dotted line is for 
           $f = 0.02$, the dashed line is for $f = 0.05$ and the 
           dot-dash line is for $f = 0.1$.}
  \label{fig:renorm_limbs}
\end{figure}

To recast the differences in Figure~\ref{fig:ldg6} into a form that is 
suitable for observational application, we define the location of the 
stellar limb to be where the surface brightness has fallen to a 
specified fraction of the central brightness, 
\begin{equation} \label{eq:rlimb}
R_\mathrm{limb} \equiv r \ \mathrm{where} \ \frac{I(r)}
                                                 {I(r = 0)} = f.
\end{equation}
Using this definition of $R_\mathrm{limb}$, we renormalize the surface 
brightness distribution.  Figure~\ref{fig:renorm_limbs} shows the 
result of setting $f = 0.01, \ 0.02, \ 0.05 \ \mathrm{and} \ 0.1$ 
for one wavelength of one spherical model.  The wavelength 
$\lambda = 705$~nm is approximately in the middle of the range shown in 
Figure~\ref{fig:ldg6}. 
From an examination of Figure~\ref{fig:renorm_limbs} we elect to define 
$R_\mathrm{limb}$ using $f = 0.01$.  This choice gives a value of 
$R_\mathrm{limb}$ that is close to $R_\star$ while also being within
reach of observations, but clearly another value of $f$ could be chosen.
Figure~\ref{fig:ldg6} shows that the surface brightness falls off 
more steeply at the limb as the model's mass increases, which reduces 
the difference between $R_\mathrm{limb}$ and $R_\star$.  
With an angular resolution $\approx 1\%$ of the stellar disk and a 
photometric precision of $\approx 1\%$, 
observations of the surface-brightness distribution from optical 
interferometry or a planetary transit could determine this location of 
the stellar limb.

\begin{figure}
  \plotone{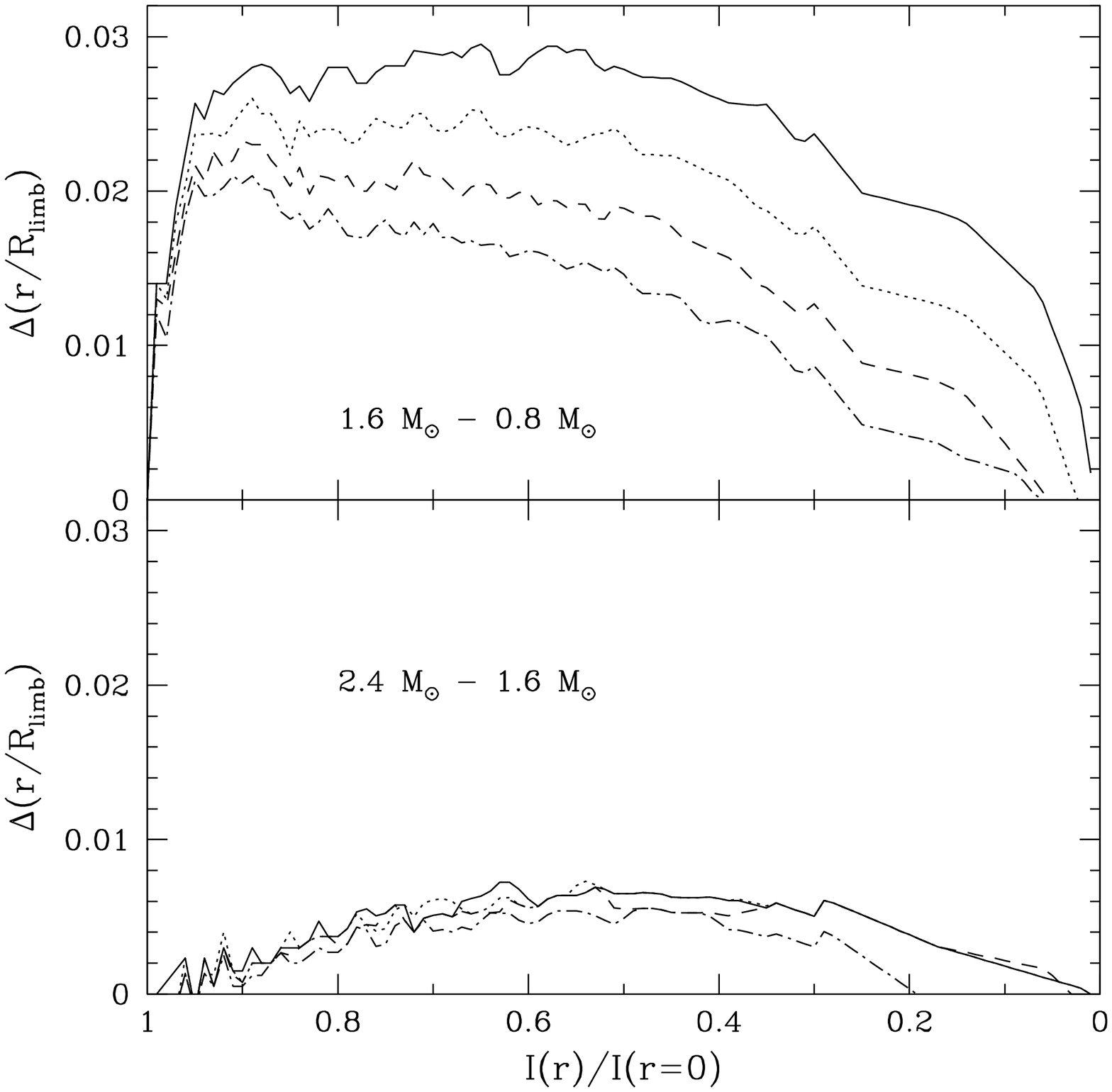}
  \caption{Difference between $r/R_\mathrm{limb}$ at $\lambda = 705$~nm
           plotted as a function of the renormalized surface-brightness 
           distribution.  The solid line is for the limb defined using 
           Equation~\ref{eq:rlimb} with $f = 0.01$, the dotted line is 
           for $f = 0.02$, the dashed line is for $f = 0.05$ and the 
           dot-dash line is for $f = 0.1$.  The top plot shows the 
           difference between the $0.8~M_{\sun}$ and $1.6~M_{\sun}$
           models, and the bottom plot shows the difference between the 
           $1.6~M_{\sun}$ and $2.4~M_{\sun}$ models.}
  \label{fig:rdiffg705}
\end{figure}

Using this observationally motivated definition of the stellar limb, we 
renormalized the intensity curves at each ODF wavelength and searched 
for the fractional radii where stellar mass produced the largest 
intensity difference.  
As an example, Figure~\ref{fig:rdiffg705} shows the differences between 
the $r/R_\mathrm{limb}$ values for the red giant models with 
$0.8~M_{\sun}$ and $1.6~M_{\sun}$ and between $1.6~M_{\sun}$ and 
$2.4~M_{\sun}$ models for the various limb renormalizations considered 
in Figure~\ref{fig:renorm_limbs}.  
The wavelength $\lambda = 705$~nm is typical of a wide range of 
wavelengths.  The difference between the $1.6~M_{\sun}$ and 
$2.4~M_{\sun}$ models is nearly the same for all the limb 
renormalizations, with the difference being greatest near 
$I(r)/I(r=0) \approx 0.5$.  The difference between the $0.8~M_{\sun}$ 
and $1.6~M_{\sun}$ models is also greatest near 
$I(r)/I(r=0) \approx 0.5$ for the renormalization at $f = 0.01$, but as 
the renormalization point is located farther in from $R_\star$, the 
location of the greatest difference between the two models also moves 
inward.  We elect to use the difference at $I(r)/I(r=0) =0.5$ because 
it applies to the limb normalization with $f = 0.01$ and because the 
half-intensity point is easy to determine observationally.

\begin{figure}
  \plotone{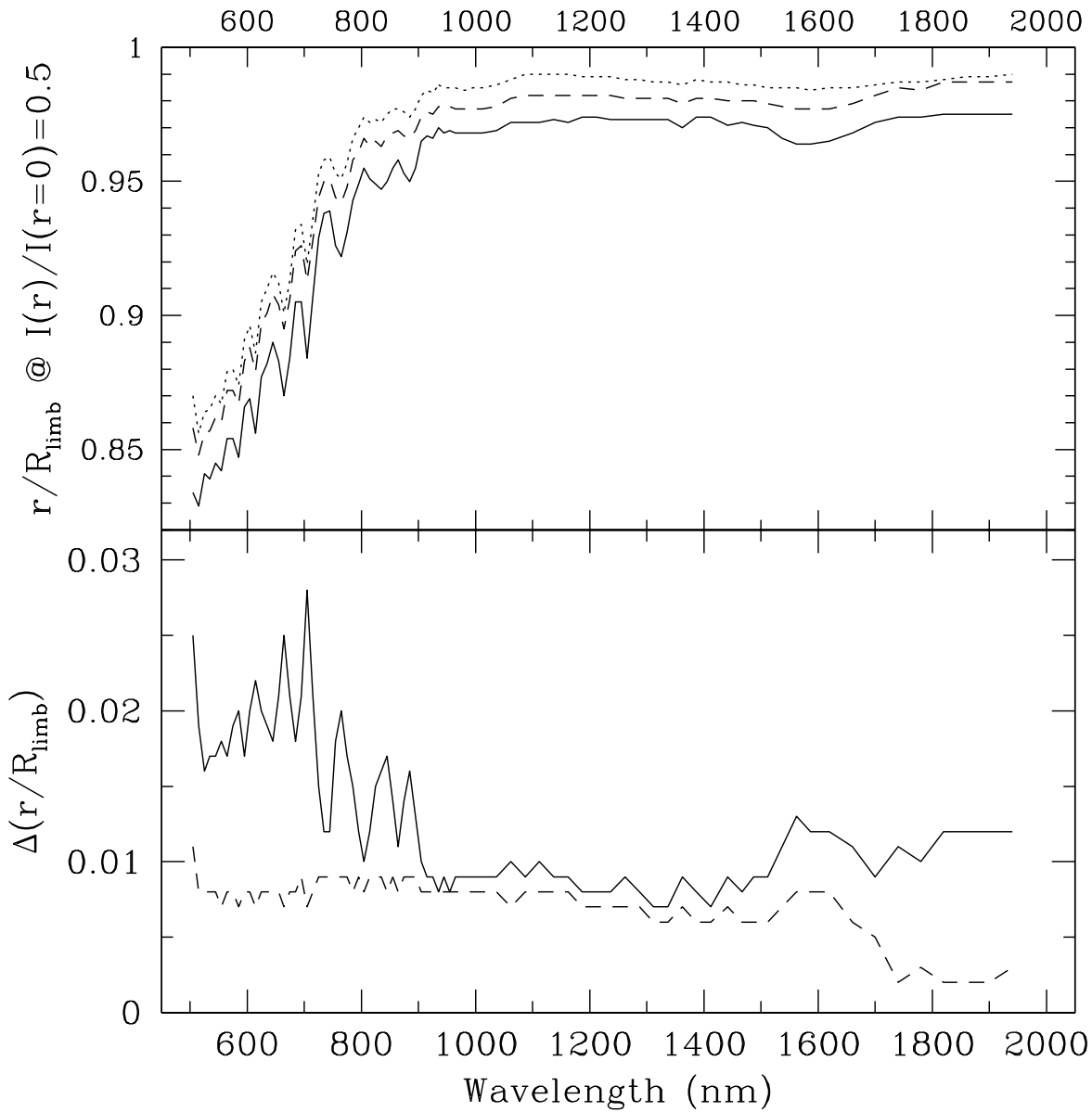}
  \caption{Top plot displays for the three red giant spherical models 
           the wavelength dependence of the renormalized fractional 
           radius ($r/R_\mathrm{limb}$) where the surface brightness is 
           half the central brightness.  The spherical models all have 
           $L_\star = 500~L_{\sun}$ and $R_\star = 50~R_{\sun}$. The 
           solid line is for $M_\star = 0.8~M_{\sun}$, the dashed line 
           is for $M_\star = 1.6~M_{\sun}$ and the dotted line is for 
           $M_\star = 2.4~M_{\sun}$.
           Bottom plot is the differences between the models in the top 
           panel, with the solid line being the difference between the 
           $0.8~M_{\sun}$ and $1.6~M_{\sun}$ models and the dashed line 
           is the difference between the $1.6~M_{\sun}$ and 
           $2.4~M_{\sun}$ models.}
  \label{fig:r05g}
\end{figure}

We determined the renormalized fractional stellar radius, 
$r/R_\mathrm{limb}$, at the half-intensity point for each ODF 
wavelength from 500~nm to 2000~nm, which is shown in 
Figure~\ref{fig:r05g}.
The models show similar trends with wavelength.  For $\la 900$~nm the 
half-intensity point moves toward the limb as the wavelength increases, 
but for $\lambda > 900$~nm the fractional radius of the half-intensity 
point is nearly constant.  Within these general trends the three masses 
are offset from each other, with the location of the half-intensity 
point shifting toward the limb as the mass increases.  It is also clear 
from the bottom panel of Figure~\ref{fig:r05g} that the location of 
the half-intensity point varies significantly with wavelength for the 
lowest mass model considered here.  For example, the separation is 
relatively larger at 705~nm, the wavelength used in 
Figure~\ref{fig:renorm_limbs}, than at adjacent wavelengths.  The 
separation also increases near 1600~nm, where the continuous opacity 
due to H$^-$ is minimum.  

Examining the stellar spectrum in the regions 
of larger separation, we find that they have relatively \emph{less} 
line opacity.  Because of this, the intensity at these wavelengths 
comes from the thermal emission deeper in the atmosphere.  As the mass 
of the star increases, the greater gas density produces a larger amount 
of thermal emission.  As a result, the surface-brightness distribution 
of the more massive star is flatter, and its half-intensity point 
shifts toward the limb.  The conclusion is that it is important to 
measure the surface-brightness distribution with sufficient spectral 
resolution, $R \approx 50-100$, to isolate the most sensitive spectral 
windows.  Of the existing interferometers capable of working in the 
optical region where the separation between the curves is greatest, the 
CHARA array's VEGA instrument \citep{Mourard2009}, which observes the 
band from 0.45 to 0.85 $\mu$m, has three choices of spectral resolution,
$R =$ 1700, 6000 and 30000, all of which are substantially greater than 
the required spectral resolving power.  The Navy Precision Optical 
Interferometer \citep{Armstrong1998} disperses the spectral band from 
450 to 850~nm onto 32 channels, providing $R \sim 50$, just at the 
lower edge of the required spectral resolution.
In the infrared, the AMBER instrument on the VLTI \citep{Petrov2003} 
could easily resolve the H$^-$ feature.  The bigger challenge for all 
the interferometers is achieving the angular resolution needed to 
locate the stellar disk's half-intensity point with an uncertainty of 
just a few percent.  Although this appears to be beyond the current 
capability, the rapid progress being made in this active field makes 
the needed improvement seem likely.

\subsection{Spectrum} \label{subsec:sec2_2}

Because the model's mass affects its structure, there can be 
associated variations of the model's spectral flux.  To test this, we 
computed synthetic flux spectra covering the wavelength range from 400 
to 1000~nm using the red giant spherical model atmospheres with 
$M_\star = 0.8~M_{\sun}$ and $2.4~M_{\sun}$.  These spectra, which were 
computed using spherical radiative transfer with the spherical model 
structure, have a spectral resolving power of $R = 10^5$ and include 
the atomic, ionic and molecular lines from the lists on Kurucz's web 
site (http://kurucz.harvard.edu/linelists.html).  Because we are 
comparing models to each other, not attempting to match observations, 
we have not included rotation or various kinds of broadening beyond the 
depth-independent microturbulent velocity of 2 km/s that is used in 
the ODFs to compute the model atmospheres.

From a comparison of the synthetic spectra, we found that $\approx 95$\%
of the lines either remained nearly constant or became weaker as the 
stellar mass increased.  For the largest weakening, which occurred in 
the 400-500~nm wavelength band, the line core decreased by about 0.15 
in residual intensity for the mass range of our models.  The amount of 
line weakening diminished steadily with increasing wavelength, with our 
models in the 900-1000~nm interval having line core residual depths 
that were smaller by no more than about 0.07.  

\begin{figure}
  \plotone{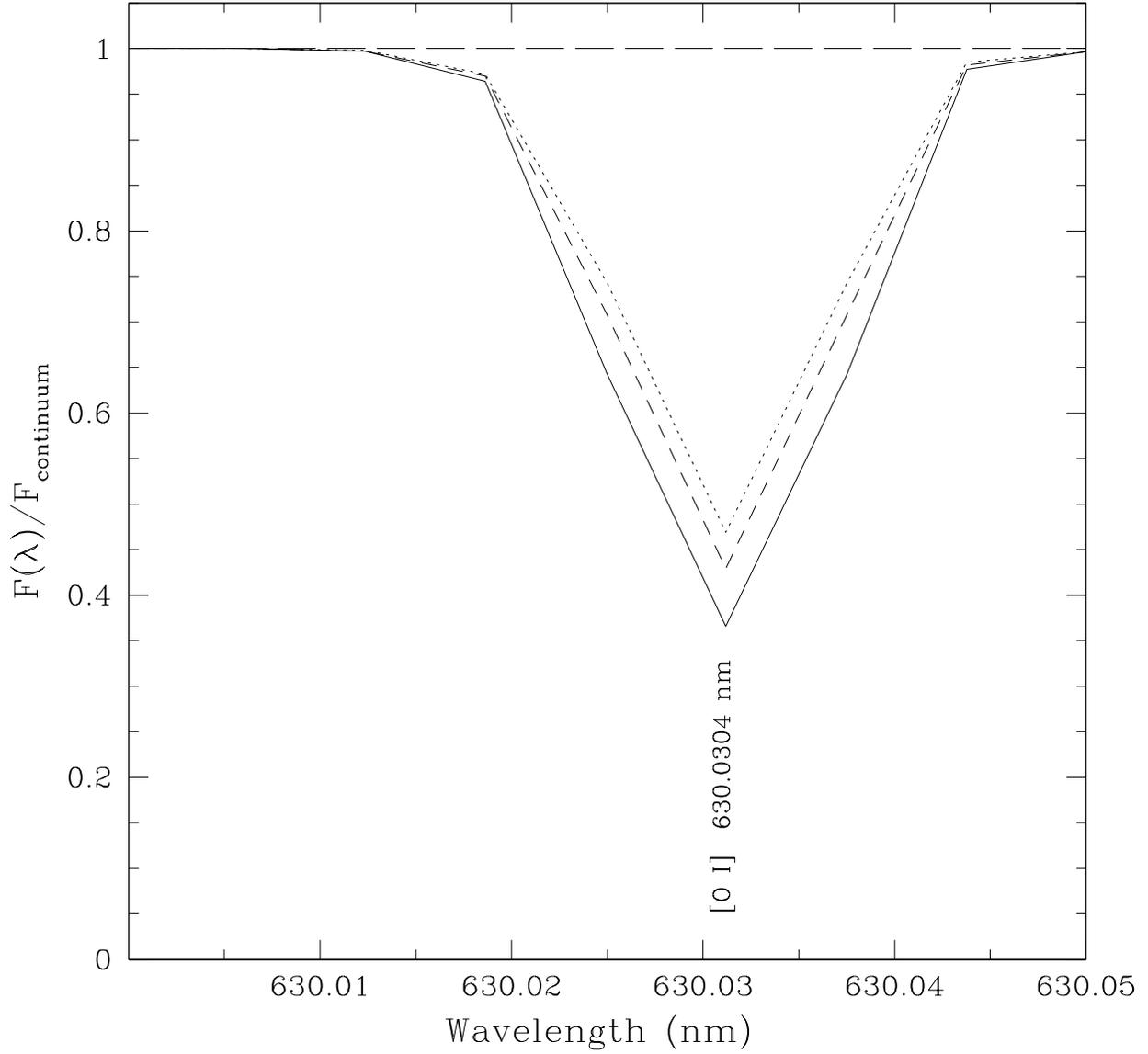}
  \caption{Rectified flux spectra of the line [O~I] 
           $\lambda =$ 630.0304~nm for the red giant spherical models 
           having $L_\star = 500~L_{\sun}$ and $R_\star = 50~R_{\sun}$. 
           The solid line is for $M_\star = 0.8~M_{\sun}$, the dashed 
           line is for $M_\star = 1.6~M_{\sun}$ and the dotted line is 
           for $M_\star = 2.4~M_{\sun}$.}
  \label{fig:o630g}
\end{figure}

At all wavelengths, the lines showing the greatest decrease with 
increasing mass were overwhelmingly ions of heavy elements, such as 
Nd~II 536.1165~nm and La~II 580.8313~nm.  However, there is one 
interesting exception to this dominance by heavy ions, which is 
[O~I] $^3$P~2p$^4 \rightarrow \, ^1$D~2p$^4$ with lines at 630.0304~nm 
and 636.3776~nm.  Figure~\ref{fig:o630g} shows the behavior of the 
630.0304~nm line, which is consistent with the work of 
\citet{Bonnell1993a} who found that [O~I] is a useful indicator of 
surface gravity.  Of course, $\log_{10} (g)$ combines $M_\star$ and 
$R_\star$, so it is a composite indicator, unlike our approach. For the 
2.4~$M_{\sun}$ star the equivalent width of [O~I] 630.0304~nm is 23\% 
less than for the 0.8~$M_{\sun}$ star, and the decrease of the 
[O~I] 636.3776~nm line is 31\%.  As shown by \citet{Bonnell1993a}, 
comparing measurements of these [O~I] lines with lines of the OH 
molecule near 1.6~$\mu$m yields the oxygen abundance and the surface 
gravity, or now the $M_\star$.

\begin{figure}
  \plotone{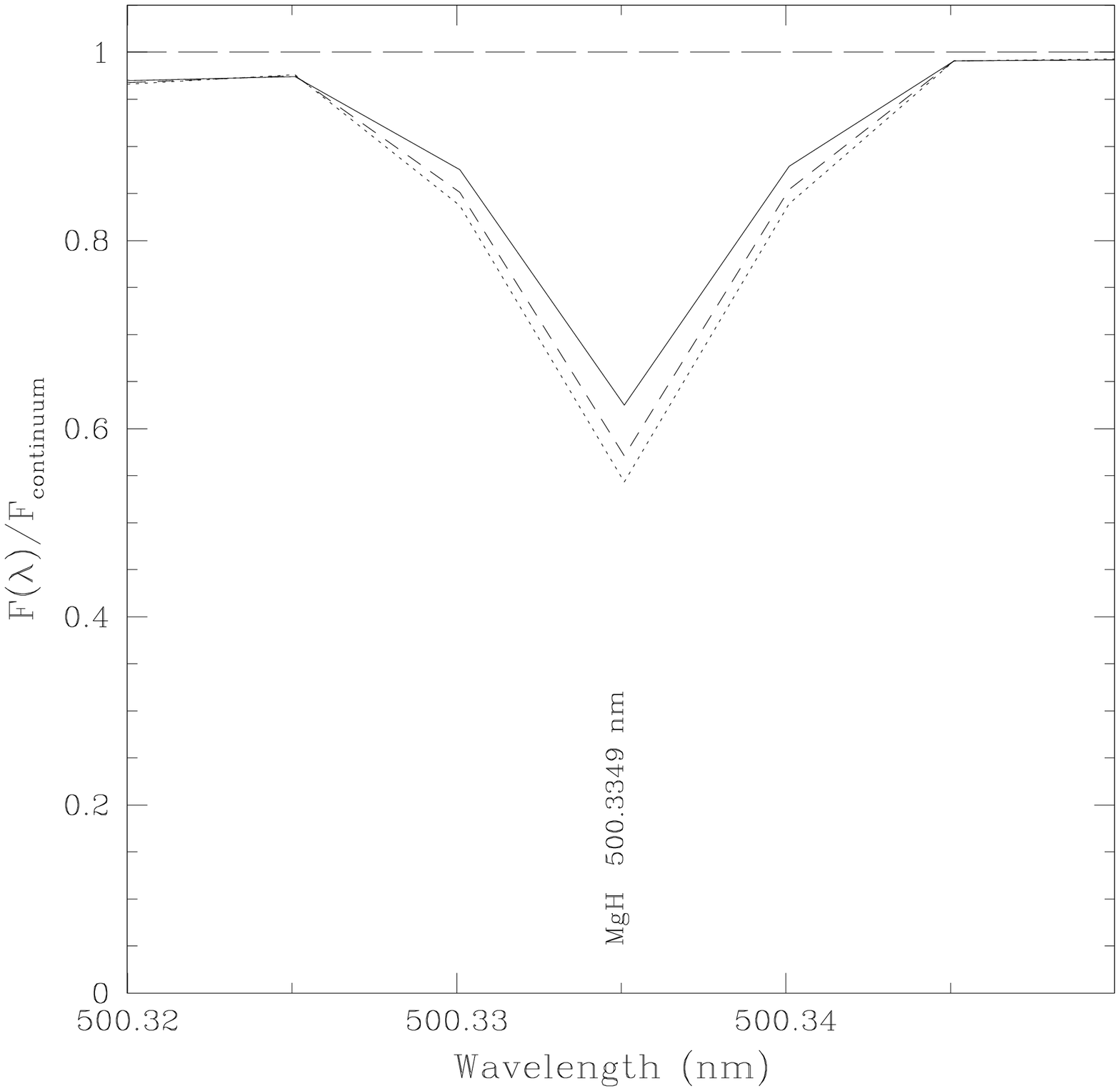}
  \caption{Rectified flux spectra of the MgH line at 
           $\lambda =$ 500.3349~nm for the red giant spherical models 
           having $L_\star = 500~L_{\sun}$ and $R_\star = 50~R_{\sun}$. 
           The solid line is for $M_\star = 0.8~M_{\sun}$, the dashed 
           line is for $M_\star = 1.6~M_{\sun}$ and the dotted line is 
           for $M_\star = 2.4~M_{\sun}$.}
  \label{fig:mgh500g}
\end{figure}

Of the $\approx 5$\% of the lines that became stronger with increasing 
mass, all of them were either MgH or SiH, with MgH strengthening by 
about twice as much as SiH.  The largest increases in the residual line 
depth, about 0.08 for our models, were for lines of MgH in the 
$X~^2\Sigma^+ \rightarrow A~^2\Pi$ band head around 500~nm.  An 
example of one line from this band is shown in Figure~\ref{fig:mgh500g},
where the line's equivalent width has become about 25\% stronger 
going from the least to most massive model.

The lines of MgH in the 500~nm spectral region have a long history of 
serving as gravity indicators, beginning with \citet{Ohman1934}.
More recently, \citet{Bell1985} and \citet{Bonnell1993b} have explored 
the use of these lines to determine the gravities of cool giants. 
By using the MgH lines in combination with lines of Mg~I, it is possible
to determine first the Mg abundance and then the surface gravity.  
Again, our focus on mass instead of the composite surface gravity 
allows for a sharper distinction.

As a test, we used the plane-parallel \textsc{Atlas\_ODF} code 
\citep{Lester2008} and the same ODF file to compute models with values 
of $T_\mathrm{eff}$ and $\log_{10} (g)$ corresponding to the red giant 
spherical models.  We used these plane-parallel models to compute 
synthetic spectra with a spectral resolving power of $10^5$ using the 
same line lists employed above.  The resulting equivalent widths of the 
[O~I] and MgH lines, and their change with surface gravity, were nearly 
identical with values computed with the spherical model.  This shows 
that these lines are sensitive to the mass, not the spherical geometry 
of the atmosphere.

\begin{figure}
  \plotone{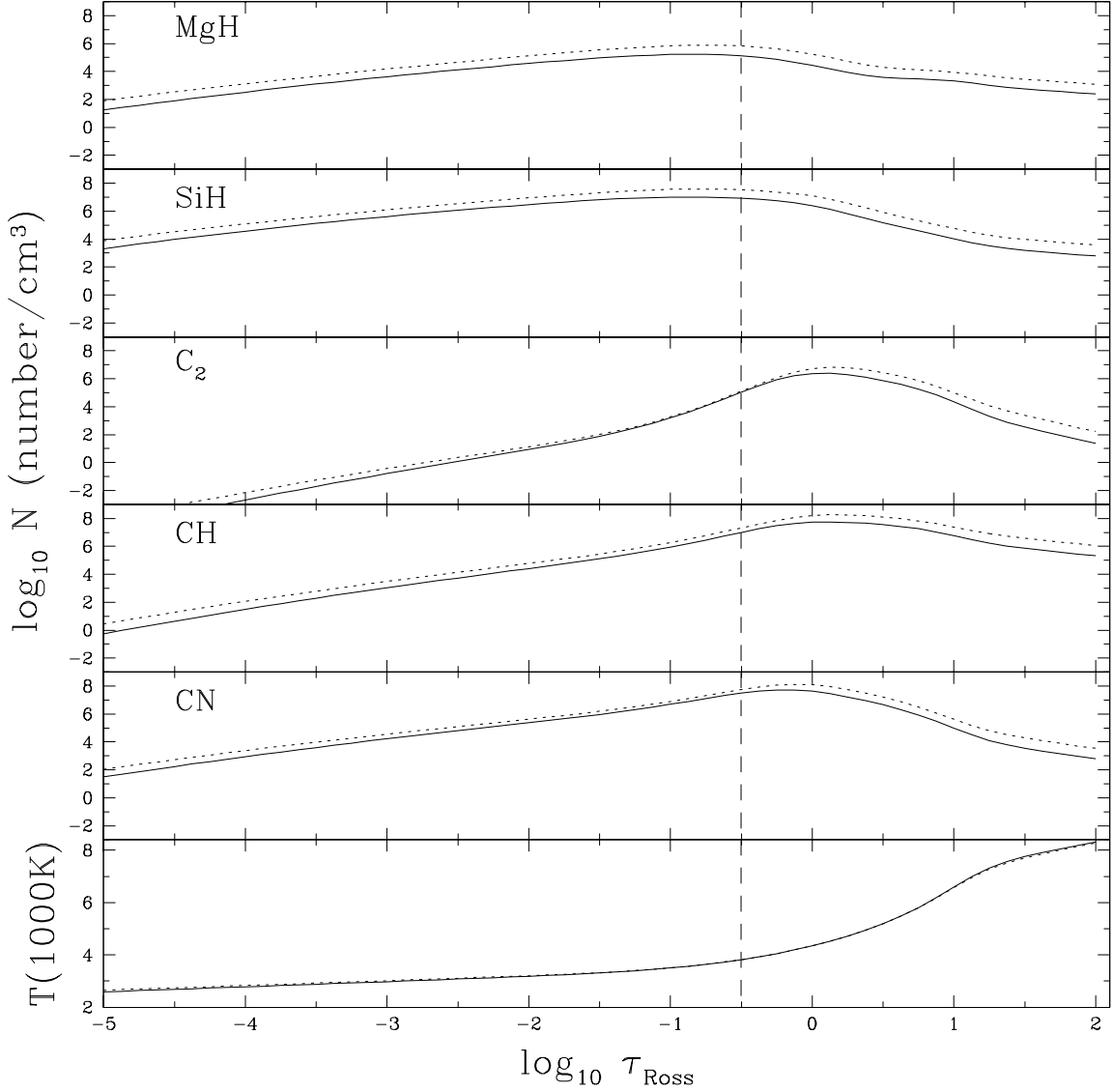}
  \caption{Top five panels show the variation of the number density of 
           prominent diatomic molecules with atmospheric depth, 
           represented by $\tau_\mathrm{Ross}$, for two red giant 
           spherical atmospheres, both having $L_\star = 500~L_{\sun}$ 
           and $R_\star = 50~R_{\sun}$.  The solid lines are for 
           $M_\star = 0.8~M_{\sun}$ and the dotted lines are for 
           $M_\star = 2.4~M_{\sun}$.  The vertical dashed line is at 
           $\log_{10} (\tau_\mathrm{Ross}) = -0.5$, which is the 
           approximate limit of our view into the stellar atmosphere.
           Bottom panel shows the variation of the temperature with 
           optical depth for the two models, which provides additional 
           context for the depth dependence of the molecular number 
           densities.}
  \label{fig:moldisg}
\end{figure}

The other molecular lines in the spectrum, except SiH, show either 
almost no change or became weaker with increasing mass.  To gain some 
perspective on this, we plot in the top panel of 
Figure~\ref{fig:moldisg} the number densities as a function of 
$\tau_\mathrm{Ross}$ for several of the prominent diatomic molecules in 
the atmospheres of the red giant models with $M_\star = 0.8~M_{\sun}$ 
and $2.4~M_{\sun}$.  In the bottom panel of Figure~\ref{fig:moldisg} we 
show the variation of the temperature with $\tau_\mathrm{Ross}$ for 
both models.
The vertical dashed line in Figure~\ref{fig:moldisg} corresponds 
approximately to the atmospheric depth where the probability of 
radiation escaping to the observer becomes negligible.  Note that only 
MgH and SiH have their peak number densities in the readily observed 
layers.  This is a consequence of these two molecules having lower 
dissociation energies than any of the other common molecules, and MgH, 
with a dissociation energy of just 1.28517~eV \citep{Shayesteh2007}, is 
more than a factor of two less than any of the other molecules.

\subsection{Composition} \label{subsec:sec2_3}

\begin{figure}
  \plotone{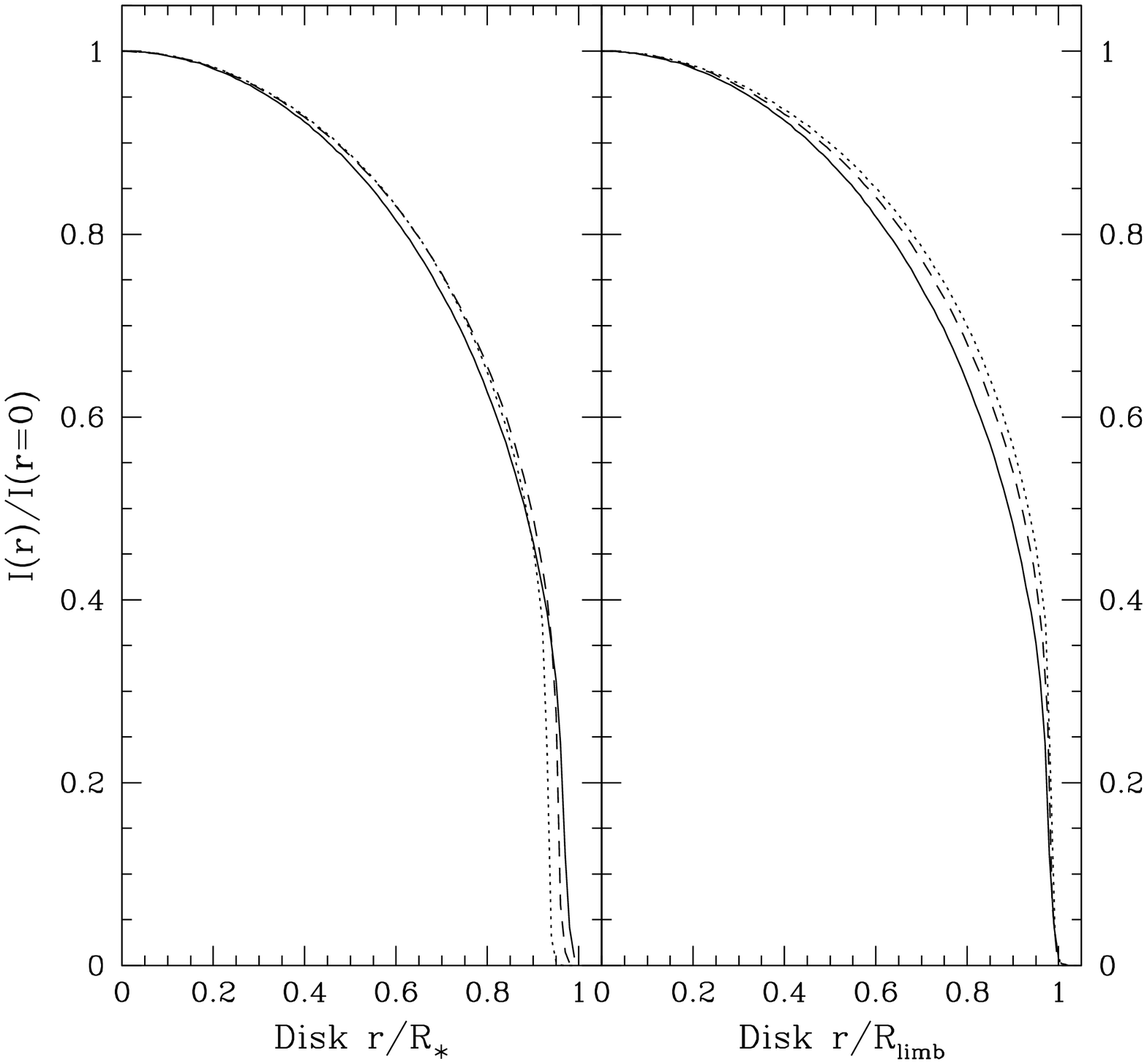}
  \caption{The left panel has the surface-brightness distributions 
           plotted as a function of the fractional disk radius, 
           $r/R_\star$, for three red giant spherical models, all 
           having the same $L_\star$, $R_\star$ and $M_\star$.  The 
           solid line is the model with solar metallicity, the dashed 
           line is the model with 0.1 solar metallicity, and the dotted 
           line is the model with 0.01 solar metallicity.
           The right panel plots the same surface-brightness 
           distributions, now as a function of the renormalized 
           fractional radius, $r/R_\mathrm{limb}$, defined by 
           equation~(\ref{eq:rlimb}) in $\S$~\ref{subsec:sec2_1}.
           The lines have the same meaning as in the left panel.}
  \label{fig:ldg_z}
\end{figure}

The spectral results discussed in $\S$~\ref{subsec:sec2_2} clearly 
depend sensitively on the composition of the star.  It is for this 
reason that it was necessary for \citet{Bonnell1993a} to use other 
spectral features to determine elemental composition before using 
either the [O~I] or MgH lines to derive surface gravity.  The 
surface-brightness distribution examined in $\S$~\ref{subsec:sec2_1} 
also depends on the metallicity of the star, which is shown in the left 
panel of Figure~\ref{fig:ldg_z}.
As the metallicity is decreased from solar, the surface-brightness 
distribution becomes fainter at the stellar limb and brighter at 
$r/R_\star \lesssim 0.9$.
Some of the relative intensities in the left panel of 
Figure~\ref{fig:ldg_z} appear to go to zero before $r$ reaches 
$R_\star$.  As explained earlier in the discussion of 
Figure~\ref{fig:ldg6}, with decreasing stellar mass the upper 
atmosphere becomes increasingly tenuous, causing I(r)/I(r=0) to drop 
below 0.01 as $r$ approaches $R_\star$.  Even though the relative 
intensity is greater than zero, it is too small to be resolved in the 
plot.

The change at the limb is obvious for the models, for which $R_\star$ 
is known, but the location of $R_\star$ is not available 
observationally.  Therefore, the same observational definition of the 
stellar limb introduced in $\S$~\ref{subsec:sec2_1} is employed here, 
with the result shown in the right panel of Figure~\ref{fig:ldg_z}.
There is now a clear progression of the surface-brightness distribution 
with metallicity, with the intensity curve steepening as the 
composition becomes more metal poor.  Therefore, to determine the 
star's mass from its intensity distribution it is essential to know 
the star's composition.  However, this requirement is much less 
stringent than what is needed to utilize a particular spectral feature, 
such as those highlighted in $\S$~\ref{subsec:sec2_2}.  Instead of 
requiring a precise value of the composition for the element producing 
a particular spectral line, only the overall composition of the star is 
needed, which can be determined by analyzing a selection of spectral 
features, or perhaps adequately estimated from spectral classification.

\section{Red Supergiants} \label{sec:sec3}

To extend our exploration of mass indicators, we next considered red 
supergiants.  Again, we use the \textsc{SAtlas\_ODF} program to compute 
atmospheric models that are spherically symmetric, this time with the 
representative parameters $L_\star = 50,000~L_{\sun}$, 
$R_\star = 500~R_{\sun}$ and solar composition.  For the masses we 
chose $M_\star = 8~M_{\sun}$, $M_\star = 16~M_{\sun}$ and 
$M_\star = 24~M_{\sun}$.  Our final models achieved constancy in the 
flux and the flux derivative very similar to the red giant models.  The 
fundamental parameters of these models yield $T_\mathrm{eff} = 3865$~K, 
the same as the red giant models, but $\log_{10} (g) = -0.06$ for the 
$8~M_{\sun}$ model, $\log_{10} (g) = 0.24$ for $16~M_{\sun}$ model and 
$\log_{10} (g) = 0.42$ for the $24~M_{\sun}$ model.  
Using the relative extension parameter defined in 
equation~(\ref{eq:eps}), the $8~M_{\sun}$ model has 
$\epsilon(-3) \approx 0.027$ and $\epsilon(-6) \approx 0.065$, the 
$16~M_{\sun}$ model has $\epsilon(-3) \approx 0.013$ and 
$\epsilon(-6) \approx 0.031$ and the $24~M_{\sun}$ model has 
$\epsilon(-3) \approx 0.008$ and $\epsilon(-6) \approx 0.020$, which, 
again, are all modest extensions.

\subsection{Surface Brightness Distribution} \label{subsec:sec3_1}

\begin{figure}
  \plotone{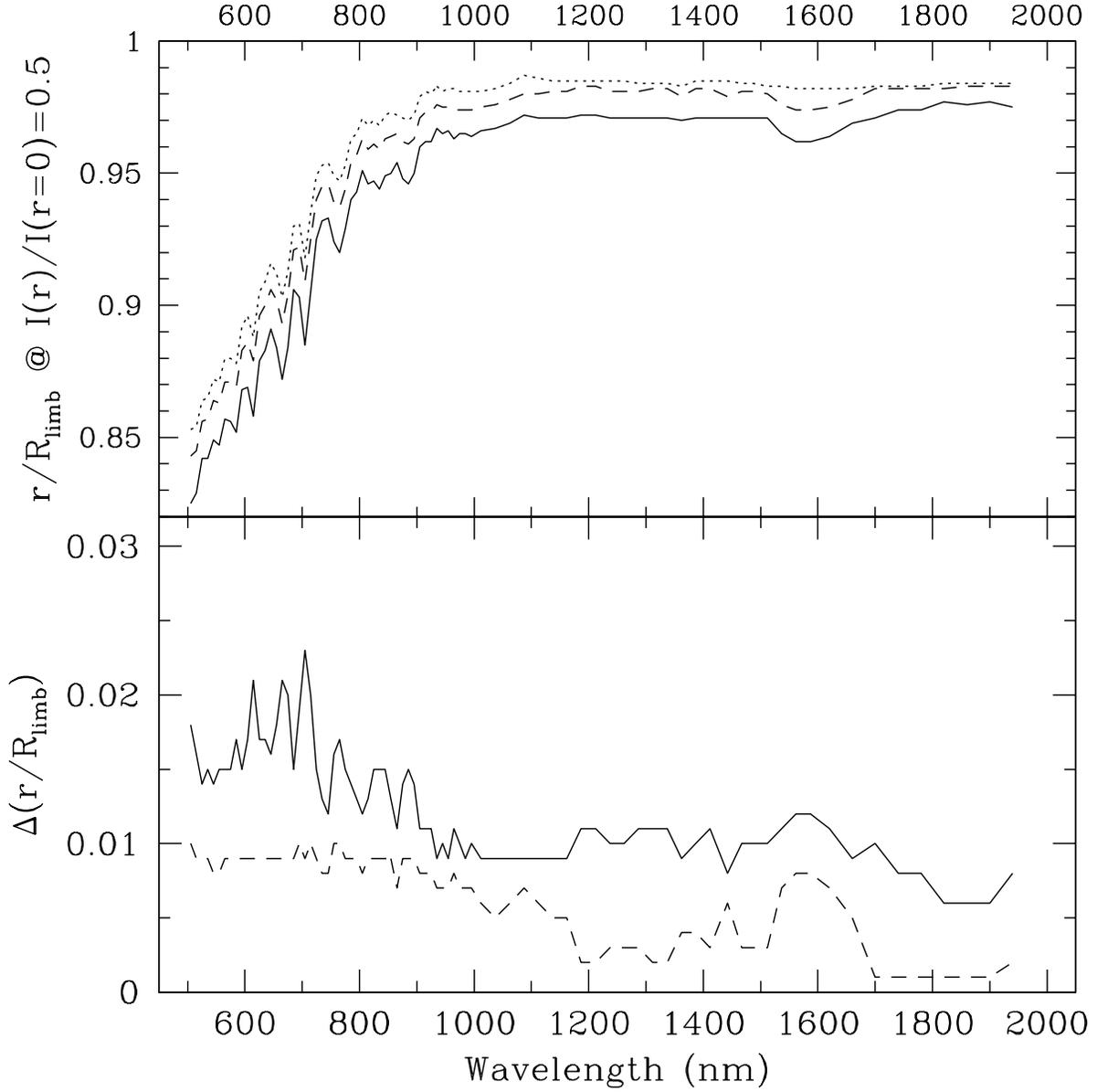}
  \caption{Top plot shows for the three red supergiant spherical models 
           the wavelength dependence of the renormalized fractional 
           radius where the surface brightness is half the central 
           brightness.  The models all have $L_\star = 50,000~L_{\sun}$,
           $R_\star = 500~R_{\sun}$ and solar composition.  The solid 
           line is the $8~M_{\sun}$ model, the dashed line is the 
           $16~M_{\sun}$ model and the dotted line is the model with 
           $M_\star = 24~M_{\sun}$.
           Bottom plot is the difference between the models in the top 
           plot, with the solid line being the difference between the 
           $8~M_{\sun}$ and $16~M_{\sun}$ models and the dashed line 
           is the difference between the $16~M_{\sun}$ and 
           $24~M_{\sun}$ models.}
  \label{fig:r05sg}
\end{figure}

The surface-brightness distributions of the red supergiants were 
computed and analyzed in exactly the same way as for the red giants, 
and the wavelength variation of the half-intensity point determined 
from the renormalized intensities is plotted in Figure~\ref{fig:r05sg}.
Comparing Figure~\ref{fig:r05sg} with Figure~\ref{fig:r05g} it is 
obvious that the trends and features are very similar, with just minor 
difference in detail.  As before, the different masses change the 
surface brightness by a few percent, and for $\lambda \la 900$~nm a 
spectral resolving power of $R \sim 50-100$ is essential to isolate 
these regions.  For $\lambda > 900$~nm the separation between the 
curves is more constant, and a lower spectral resolving power would be 
adequate to detect this difference.

\subsection{Spectrum} \label{subsec:sec3_2}

To search for signatures of mass, we used the spherical red 
supergiant model atmospheres with $M_\star = 8~M_{\sun}$ and 
$M_\star = 24~M_{\sun}$ to synthesize the flux spectrum from 400 to 
1000~nm at $R = 10^5$ using the same line lists as for the red giants.  
Comparing the line depths, we again found that $\approx 95$\% of the 
lines were either unchanged or became weaker.  As was found for the red 
giants, the lines displaying the greatest weakening with increasing 
mass were located at the shortest wavelengths we computed.  The trend 
found for the red giants, that the amount of weakening decreases with 
increasing wavelength, also holds true for the red supergiants, but the 
drop off is somewhat more gradual until the very longest wavelengths.  
The total variation over the spectral range $400 - 1000$~nm is the same.

As was found for the red giants, the lines that weaken the most with 
increasing mass were overwhelmingly ions of heavy elements, with the 
[O~I] lines remaining noteworthy exceptions.  The [O~I] lines in the 
supergiants are about twice as strong as for the giants, but their 
percentage weakening with increasing mass is somewhat less.  In 
addition, the equivalent widths of [O~I] 636.3776~nm and 630.0304~nm
for the red supergiants each weaken by comparable amounts, although the 
percentage changes are different, but for the red giants the equivalent 
width change of the 630.0304 line is almost 50\% greater than for the 
636.3776~nm line.

The $< 5$\% of the lines that become stronger with increasing mass 
are all lines of either MgH or SiH, with the MgH lines changing by 
about twice as much as the SiH lines.  Although the MgH lines of the 
supergiants are not as deep as they are in the giants, their depths 
change by about 20\% more than between the most and least massive giant 
models, and this is also true for the change of their equivalent widths.
The variation with depth of the number density of the diatomic 
molecules for the supergiant atmospheres is very similar to the 
variation with depth for the red giants shown in 
Figure~\ref{fig:moldisg}.

\section{Discussion} \label{sec:sec4}

Both the distribution of the surface brightness and the flux spectrum 
contain indicators of the mass of the red giants and supergiants.  Of 
these, the spectral features of [O~I] and MgH have been recognized and 
used previously, although they were associated with $\log_{10} (g)$, 
which depends on the combination of two of the fundamental stellar 
parameters.  With the ability to measure independently both $L_\star$ 
and $R_\star$, these lines can now be used to determine the mass of a 
non-binary star.  Current spectroscopic capability is clearly 
sufficient to measure these lines with the precision required to do the 
analysis.

The major complication in using spectral features to determine the 
stellar mass is the line variability exhibited by some giants and by 
many supergiants.  In an extended observational study of Betelgeuse 
($\alpha$ Ori, HR 2061, HD 39801) \citet{Gray2000, Gray2001, Gray2008}
recorded variations in the strength of several neutral metal lines in 
a relatively small spectral window centered on $\lambda$ 625.0~nm.
These variations were attributed to changes in the continuous 
opacity that alters the contrast between the line core and the 
adjacent continuum.  Three-dimensional radiative hydrodynamic modeling 
also predicts time variation of the spectrum due to surface 
inhomogeneities \citep{Chiavassa2011}.  The spectral analysis presented 
here found that most lines, including the neutral lines studied by 
Gray, have a negligible variation with mass.  The specific spectral 
lines that we identified probably do show the spectral variations 
caused by surface structure in addition to the dependence on mass.  
Therefore, to employ the mass-sensitive spectroscopic features found 
here, it clearly is necessary to normalize the line strengths to 
adjacent lines insensitive to mass in order to remove the variability 
caused by processes occurring within the atmosphere.  This is in 
addition to eliminating the effect of the star's composition by 
normalizing to other lines of the same element, as was done by 
\citet{Bell1985}, \citet{Bonnell1993b} and \citet{Bonnell1993a}.

The signature of mass in the surface-brightness distribution is just a 
few percent, which is smaller than the variation of the spectroscopic 
features.  However, this small signature may not be as challenging as 
it appears from our analysis.  In our tests the giant and supergiant 
masses were each varied by a factor of three, resulting in 
$\Delta \log_{10} (g) = 0.48$ in both cases, with the corresponding 
extension parameter $\epsilon(-3) \lesssim 0.03$ and 
$\epsilon(-6) \lesssim -0.06$ .  Even careful, detailed observational 
analyses of red giants and supergiants frequently have uncertainties 
$\approx 0.5$ or more in their determinations of $\log_{10} (g)$.  
Arcturus provides an illustrative example because it has data of the 
highest quality, which have been analyzed repeatedly.  The thorough 
study by \citet{Peterson1993} concluded that 
$\log_{10} (g) = 1.5 \pm 0.15$, found by minimizing the differences 
between the observed high resolution flux spectrum and spectra 
synthesized from a grid of plane-parallel \textsc{Atlas} stellar 
atmospheres.  The equally thorough examination of Arcturus by 
\citet{Griffin1999} determined that $\log_{10} (g) = 1.94 \pm 0.05$ by 
minimizing the differences between observed and computed spectral 
energy distributions, again using plane-parallel \textsc{Atlas} models. 
Recently, \citet{Ramirez2011} determined 
$\log_{10} (g) = 1.66 \pm 0.05$ for Arcturus using evolutionary 
isochrones.  The agreement of the $\log_{10} (g)$ values is just 
marginal considering the error bars.  Stars that are less well observed 
than Arcturus often have significantly larger uncertainty in their 
$\log_{10} (g)$ determinations.  If the surface-brightness distribution 
could determine $\log_{10} (g)$ with an uncertainty of 0.5, it would be 
competitive with the currently employed methods.  In addition, the 
physics responsible for the surface-brightness signature of mass is 
particularly simple, just hydrostatic and thermal equilibrium.  We did 
find that the intensity distribution depends on the general metallicity 
of the stellar atmosphere, but estimates from spectral classification 
or a more quantitative determination from a selection of spectral lines 
would be adequate to determine this parameter.  The resulting mass 
determination will provide a different perspective from the other 
methods that rely on significantly different and more complex physics.

The measurement of stellar surface brightness distributions is a 
revolutionary observational achievement, but one that is still 
developing.  For example, the recent study by \citet{Lacour2008} of 
Arcturus, a star whose disk is well resolved by optical/infrared 
interferometry, concluded that the disk was not uniformly bright, but 
was unable to observe exactly how the surface brightness varied with 
position on the disk.  Several studies of Betelgeuse 
\citep{Buscher1990, Tuthill1997, Young2000, Haubois2009} have detected 
large bright features on its surface, while other studies 
\citep{Ohnaka2009, Ohnaka2011} have found only slight deviation from a 
uniform disk, significantly smoother than the predictions of the 
three-dimensional radiative hydrodynamic models \citep{Chiavassa2009}.
However, when observational agreement is reached about the true 
appearance of the surfaces of these stars, a key point is that the 
features will be due to physical processes taking place within the 
atmosphere.  These disk substructures will complicate the observations, 
but they do not invalidate the conclusion reached here that the basic 
trend of the stellar surface brightness distribution contains the 
signature of the fundamental global stellar parameter of mass.

\citet{Neilson2011, Neilson2012} have recently explored a different, 
more statistical way of using the surface-brightness distribution of a 
cool, luminous star to determine its mass.  Using the same 
\textsc{SAtlas\_ODF} code, they computed thousands of spherical model 
atmospheres with $L_\star$, $M_\star$ and $R_\star$ thoroughly covering 
the parameter space of the red giants and supergiants.  After computing 
the surface-brightness distribution for each model, a limb-darkening 
law with two coefficients was fit to the intensities.  An analysis of 
the ensemble's limb-darkening coefficients showed that they depend on 
the atmospheric extension, characterized by the parameter 
$R_\star/M_\star$.  By fitting the limb-darkening law to a star's 
observed surface-brightness distribution, the coefficients determine 
the star's $R_\star/M_\star$.  Determining the star's $R_\star$ using 
optical interferometry and parallax measurements gives the star's mass. 
The approach taken by \citet{Neilson2011, Neilson2012} is quite 
different from the analysis done here, but the conclusion is the same; 
it is possible to determine the mass of a cool, luminous star by 
analyzing its surface brightness distribution.

A potential interesting variation of the methods explored here is to 
apply them to radially pulsating stars, such as Cepheids.  Even though 
Cepheids lose mass over time via a stellar wind \citep{Neilson2008}, at 
a given epoch the Cepheid's mass is constant.  By measuring the 
periodic variation of the star's luminosity and radius, the diagnostics 
explored here would be able to determine the Cepheid's mass at its 
current evolutionary stage.

\section{Conclusions} \label{sec:sec5}

We have used the \textsc{SAtlas\_ODF} program \citep{Lester2008} to compute 
spherical model stellar atmospheres characterized by the three 
fundamental parameters $L_\star$, $M_\star$ and $R_\star$, which were 
chosen to represent typical red giants and supergiants.  Because it is 
possible to observe $L_\star$ and $R_\star$, we held the luminosity and 
radius constant and varied the mass.  Searching the radiation of the 
models for features that varied with $M_\star$, we found that there is 
a systematic change in the surface-brightness distribution, which we 
characterize by the renormalized fractional radius, $r/R_\mathrm{limb}$,
at the stellar disk's half-intensity point, $I(r)/I(r=0) = 0.5$.  For 
the giant and supergiant models used here, the $r/R_\mathrm{limb}$ 
varies by a few percent for the mass range we explored, which is 
currently below observational detection.  However, with a larger mass 
variation, and a corresponding larger atmospheric extension, and with 
improving observational techniques, this signature will become viable.

We also synthesized the flux spectra of the red giant and supergiant 
spherical models with a spectral resolving power of $10^5$, taking into 
account the spherical radiative transfer.  The most interesting changes 
with mass were found to be the lines of [O~I], which become weaker with 
increasing mass, and numerous lines of MgH in the 
$X~^2\Sigma^+ \rightarrow A~^2\Pi$ band head around 500~nm, which 
strengthen with increasing mass.  Both of these spectral features have 
been used previously as indicators of $\log_{10} (g)$, and they can be 
used now to determine stellar mass with current spectral capabilities.

\acknowledgements

This work has been supported by a Discovery grant from the Natural 
Sciences and Engineering Research Council of Canada to JBL, and HN 
acknowledges funding from the Alexander von Humboldt Foundation.
We thank the referee for his/her many detailed comments, questions and 
suggestions that led us to examine our results more closely and to 
explain them more clearly.

%\bibliography{mi} % reference mi.bib

\end{document}